# An alternative interpretation of linguistic variables as Linguistic Finite Automata

**Supriya Raheja[1], Reena Dhadich[2] and Smita Rajpal[1]**

[1] Department of CSE, ITM University
Gurgaon, Haryana, India
*Supriya.raheja@gmail.com, smita_rajpal@yahoo.co.in*

[2] Department of Computer Science, Banasthali University
Banasthali, Rajasthan, India
*reena.dadhich@gmail.com*

**Abstract**
Linguistic variables represent crisp information in a form and precision appropriate for the problem. For example, to answer the question "How are you?" one may say "I am fine." the linguistic variables like "fine", so common in everyday speech. In this paper an alternative interpretation of linguistic variables is introduced with the notion of a linguistic description of a value or set of values. The use of linguistic variables in many applications reduces the overall computation complexity of the application. Linguistic variables have been shown to be particularly useful in complex non-linear applications. Here we are applying the concept of reasoning with Linguistic Quantifiers to define the Linguistic Finite Automata along with the expansion of $\delta$ and $\lambda$ over $\delta$ and $\lambda$.

## 1. Introduction

The concept of a linguistic variable was first introduced by Zadeh [6] as a model of how words or labels can represent vague concepts in natural language. Some Formal Definitions we are discussing here as:

**Definition 1.1(Linguistic variable):** A linguistic variable is a quadruple $[L, T(L), \Omega, M]$ in which L is the name of the variable, $T(L)$ is a countable term set of labels or words (i.e. the linguistic values), $\Omega$ is a universe of discourse and M is a semantic rule.

The semantic rule M is defined as a function that associates a normalized fuzzy subset of X with each word in T (L). In other words the fuzzy set M (w) can be viewed as encoding the meaning of w so that for $u \in U$ the membership value $\mu_M(w)(u)$ quantifies the suitability or applicability of the word w as a label for the value u. We can regard the semantic function M as being determined by a group voting model [7] across a population of voters as follows. Each voter is asked to provide the subset of words from the finite set T (L) which are appropriate as labels for the value u. The membership value $\mu_M(w)(u)$ is then taken to be the proportion of voters who include w in their set of labels.

In [6] Zadeh originally defined a linguistic variable as a quintuple by including a syntactic rule according to which new terms (i.e., linguistic values) could be formed by applying hedges to existing words. However, the semantics of such hedges seem far from clear and the rather arbitrary definitions given in [6] appear inadequate. Indeed, in our view, it is far from apparent that there should be a simple functional relationship between the meanings of a word and the meaning of a new word generated from it by applying a hedge. In other words, we would claim that while hedges are a simple syntactic device for generating new terms there is no equally simple semantic device for generating the associated new meanings. Hence, in the following we consider only fixed finite term sets where all the labels and their associated meanings are predefined. This does not mean that we do not permit labels such as very small in the term set but rather that we would take its meaning as being predefined instead of being determined from that of small in a functional way.

*Example 1.1* Consider the set of words {small (s), medium (m), large(l)} as labels of a linguistic variable SIZE describing values in U =[ 0, 100] . Given a set of 10 voters a possible voting pattern for the value 25 is

Table 1: Voting Pattern

| Voter 1 | Voter 2 | Voter 3 | Voter 4 | Voter 5 | Voter 6 | Voter 7 | Voter 8 | Voter 9 | Voter 10 |
|---------|---------|---------|---------|---------|---------|---------|---------|---------|----------|
| small | small | small | small | small | small | small | small | small | small |
| Medium | Medium | medium | medium | medium | | | | | |

This gives $\mu_{M(small)}(25)=1$ and $\mu_{M(Medium)}(25)=0.5$

Now this voting parameter can be represented by a mass assignment on the power set of {small, medium, large},{small, medium}:0.5,{small}:0.5(i.e. 50% of voters select both small and medium as possible labels for 25 and 50% select only small). This in turn represents a fuzzy set on the set of words, namely

Small/1+medium/0.5

Hence, in practice we need only define the fuzzy sets M (small), M(medium) and M (large) from which we can determine any linguistic description. Fig. 1 illustrates how a linguistic description can be ``read'' from the fuzzy set meanings of the words. Here the value 25 has membership 1 in M (small), 0.5 in M( medium) and zero in M(large)(and all other labels) giving a linguistic description of small/1 + medium/0.5.

**Definition 1.2 (Linguistic description of a value):** Let x $\in \Omega$. Then the linguistic description of x relative to the linguistic variable L is the fuzzy subset of T (L).

$$des_L(X) = \sum_{w \in T(L)} w/\mu_{M(w)}(X) \quad (1)$$

In cases where the linguistic variable is fixed we drop the subscript L and write des(x). This notion can be extended to the case where the value given is a crisp set or a fuzzy subset of $\Omega$ in which case the appropriate linguistic description is defined as follows.

**Example 1.3:** Let X be a variable with values in $\Omega$ [0, 10] and let [L, T(L), $\Omega$, M] be a linguistic variable labeling X, where T(L) {very small, small, medium, large, very large} and where M(very small) =[ 0:1 2:1 3:0], M (small)=[1:0 2:1 4:1 5:0], M(medium)=[3:04:16:17:0], M(large)=[5:0 6: 1 8:1 9:0], M(very large=[7:0 8:1 10:1].

Here we are using the notation for piecewise linear function where [$x_1:y_1,\ldots x_n: y_n$] denotes a function F(x) such that $\forall x \in [x_i,\ldots x_i+1]$.
F(x) = ($y_i-y_i+1/x_i-x_i+1$) + ($x_iy_i+1-y_ix_i+1/ x_i-x_i+1$) for i =1… n.

Now let us define the linguistic description of the set

Pr (Medium|$des_L(x)$) = x-3/2 for x $\in$ [3,5]= 7-x/2 for x $\in$ [5,7]

**Example 1.2**: Consider a linguistic variable $\langle$ SALARY, {low, moderate, good, very good},[15; 50] , M $\rangle$ labeling the salary of project managers in India. The salary values are in thousands. The semantic function M is defined such that:

M (low) = [15:1, 20:1, 25:0]
M (Moderate) = [17.5: 0; 20: 1, 30: 1, 32:5: 0]
M (good) = [25: 0, 30: 1, 35: 1, 42:5: 0]
M (very good) = [32.5:0, 35:1, 50:1]

Now suppose for a certain company the following linguistic description has been obtained for the salary of consultant employed there

Des=low/0.33+moderate/0.95+good/1+verygood/0.44.

This has mass assignment

{low, moderate, good, very good}:0.33,
{moderate, good, very good} : 0:11,
{moderate; good} : 0:51,
{good}: 0:05.

Hence, the density function (1) on salary given this information is

P(x|des)=0.33p( x| low , moderate, good, very good)
    +0.11p(x|moderate, good, very good)
    +0.51p(x| moderate, good)+0.05 p ( x| good)

## 2. Representing expressions with linguistic quantifiers

In natural language linguistic quantifiers are used to indicate the speaker's level of belief in a statement or to express the degree to which it is applicable.

For instance, the following are typical English sentences:

Most good musicians like dance,

It is highly likely that India will win the world cup.

In both these cases we can interpret the quantifier as a linguistic description of the probability of the statement. That is the quantifiers are words or labels describing a probability value and their meanings are given by fuzzy subsets of the interval [0, 1]. In the case of the first of the two sentences above we might argue that it is more natural to think of most as describing the proportion of the set of all mathematicians who like music. However, clearly we can equally think of such a proportion as the probability of picking a mathematician at random who likes music. It should also be noted that we restrict linguistic quantifiers to descriptions of probabilistic belief values and do not allow quantification over fuzzy truth values as illustrated by terms such as quite true and very true [8]. It is our view, as is consistent with the voting model, that truth values are emergent properties and can only be defined across a population of voters. Hence, it is not meaningful to discuss truth values with individual voters and in particular it is not meaningful to ask individual voters to provide labels for truth values. We would also suggest that this position is in keeping with natural language usage since, at least in English, quantifiers such as fairly true rarely occur and when they do, such as in the response ``Yes, that's very true'', they merely serve to emphasize a binary truth value.

In this section we describe a calculus for reasoning with linguistically quantified expressions of the form described above. Initially, however, we introduce a language consisting of formula and conditional formula involving linguistic variables.

The symbol of the language consist of a set of linguistic variables $L_1,\ldots,L_k$. The Logical Connectives $\neg$, $\wedge$, $\vee$ and a conditional divider |.The constants are the element of $T(L_i)$ for i=1,……..,k. Atomic Linguistic formula have the form(L=w) or $\neg$ (L=w),where L is the Linguistic variable and w$\in$T(L).The well formed formula of this language is defined recursively as follows:

**Definition 2.1 Linguistic Formula:**
All atomic linguistic formula are linguistic formula. Furthermore, if F1 and F2 are linguistic formula, then so are:
1. ($\neg$ F1) and ($\neg$ F2),
2. (F1 $\wedge$ F2),
3. (F1 $\vee$ F2)

**Definition 2.2 Conditional linguistic formula:**
The expression (H | B) is a conditional linguistic formula if H is an atomic linguistic formula and B is a linguistic formula.

**Definition 2.3 A quantifier variable:**
A quantifier variable is a linguistic variable ($L_a$, QS, [0, 1], M), where a is the probability of some linguistic formula and QS is a finite totally ordered set of words $\{Q_1,...,Q_n\}$ with meanings forming a linguistic covering of [0, 1] and such that $\forall$ x $\in [0,1] \mu_M(Q_i)$ (x)= $\mu_M(Q_{n-i}+1)(1-x)$. In this case $Q_{n-i}+1$ is said to be the antonym of $Q_i$, denoted ant($Q_i$), and vice versa. It is generally assumed that QS and M are fixed across all quantifier variables.

## 3. Application of Linguistic variables in Finite Automata

3.1 Linguistic Finite Automata

Consider a linguistic Finite Automaton defined as five tuples as $\langle Q, \Sigma, \delta, q0, F \rangle$, where Q is the finite Set of States, $\Sigma$ is the finite set of input alphabets, $\delta$ is the transition Function, q0 is the initial state and F is the final accepting state. Where $\delta$ is defined as Q X $\Sigma$ into Q.

$\delta$ : Q × $\Sigma$ × Q → [0, 1]

($q_i$, xs, $q_j$) → ($q_i$, xs, $q_j$) = $\theta$ ij(xs);

$\delta$ (xs) = [$\theta$ ij(xs)]n× n; s= 1, 2, ,…, m.

$\lambda$ : Q × $\Sigma$ → [0, 1]

(qi, xs, yt) → $\lambda$ (qi, xs, yt) = $\pi$ it(xs);

$\lambda$ (xs) = [$\pi$ it(xs)]n× l; s= 1, 2, … , m.

The following conditions hold:
(1) $\forall$q $\in$Q, x $\in \Sigma$ $\exists$ p $\in$Q, such that $\delta$ (q, x, p)>0 $\Rightarrow$ $\exists$y $\in \Delta$, such that $\lambda$ (q, x, y)>0.
(2) $\forall$q $\in$Q, x $\in \Sigma$ , $\exists$y $\in \Delta$, such that $\lambda$ (q, x, y)>0$\Rightarrow \exists$p $\in$ Q, such that $\delta$ (q, x, p)>0.

Let $\Sigma^*$ denote the set of all words of finite length over $\Sigma$ and $\Delta^*$ denote the set of all words of finite length over $\Delta$. Let ε denotes the empty word. For x $\in \Sigma^*$ and y $\in \Delta^*$, |x| denotes the length of x and |y| denotes the length of y.
In the case of multi-input sequence $xs_1$, $xs_2 \cdot \cdot \cdot xs_m$.

$$\delta': Q \times \Sigma^* \times Q \to [0,1] \quad (2)$$

$(q_i, \varepsilon, q_j) \to \delta^\square(q_i, \varepsilon, q_j) = 1$ if $q_i = q_j$,
$\qquad\qquad\qquad\qquad = 0$ if $q_i \neq q_j$ (2).

$(q_i, xs_1\ xs_2 \cdots xs_m, q_j) \to \delta^\square(q_i, xs_1\ xs_2 \cdots xs_m, q_j)$

$= \max\{\min[(\delta(q_i, xs_1, q_o); \delta(q_o, xs_2, q_p), \ldots \delta(q_v, xs_m, q_j)]\}, q_o, q_p \ldots q_v \in Q$

$= \max\{\min[\theta_{io}(xs_1); \theta_{op}(xs_2), \ldots, \theta_{vj}(xs_{m\_})]\}$
$\quad q_o, q_p \ldots q_v \in Q$

$\delta^\square(\varepsilon) = E_{n \times n}$ : $E_{n \times n}$ is n-order identity matrix.

$\delta^\square(xs_1\ xs_2 \cdots xs_m) = \delta(xs_1) \circ \delta(xs_2) \circ \ldots \circ \delta(xs_m)$

$$\lambda: Q \times \Sigma^* \times \Delta^* \to [0,1] \quad (3)$$

$(q, x, y) \to \lambda^\square(q, x, y) = 1$ if $x = y = \varepsilon$.
$\qquad\qquad\qquad\qquad = 0$ if $x \neq \varepsilon, y = \varepsilon$ or $x = \varepsilon, y \neq \varepsilon$

$(q, x_a, y_b) \to \lambda^\square(q, x_a, y_b) = \vee\ \{\lambda^\square(q, x, y) \wedge \delta^\square(q, x, r) \wedge \lambda(r, a, b) | r \in Q\}$ (3)

$= \lambda^\square(q, x, y) \wedge \{\vee [\delta^\square(q, x, r) \wedge \lambda(r, a, b)| r \in Q]\}$
$q \in Q, a \in \Sigma, b \in \Delta^*, x \in \Sigma^*, y \in \Delta^*$

Compared with the old model of Finite Automata holding all the conditions our modified model Linguistic Finite Automata is able to handle the Linguistic data and transitions which is not possible for the general kind of Finite Machine.

Now we are defining the new definition of equivalence relations of Linguistic Finites Automata:

**Definition 3.1:**
Let $M_i = (Q_i, \Sigma, \Delta, \delta_i, \lambda_i)$ be a Linguistic Finite Automata with output $\Delta$ and Output Function $\lambda$ which maps Q into $\Delta$, i = 1,2. Let $q_i \in Q_i$, i = 1,2 Then the equivalence relations for q1 and q2 are defined as:

(1) $q_1$ and $q_2$ are equivalent $(q_1 \equiv q_2) \Leftrightarrow \forall x \in \Sigma^*, y \in \Delta^*, \lambda^\square_1(q_1, x, y) = \lambda^\square_2(q_2, x, y)$.

(2) For each positive integer k, $q_1$ and $q_2$ are k-equivalent $(q_1 \equiv_k q_2) \Leftrightarrow \forall x \in \Sigma^*$ of length $\leq k$, $y \in \Delta^*, \lambda^\square_1(q_1, x, y) = \lambda^\square_2(q_2, x, y)$.

(3) M1 and M2 are equivalent $(M1 \equiv M2) \Leftrightarrow \forall q1 \in Q1, \exists q2 \in Q2$, such that $q1 \equiv q2$ and $\forall q2 \in Q2, \exists q1 \in Q1$, such that $q2 \equiv q1$.

(4) For each positive integer k, $M_1$ and $M_2$ are k-equivalent $(M_1 \equiv_k M_2) \Leftrightarrow \forall q_1 \in Q_1, \exists q_2 \in Q_2$, such that $q_1 \equiv_k q_2$ and $\forall q_2 \in Q_2, \exists q_1 \in Q_1$, such that $q_2 \equiv_k q_1$.

If $M_1 = M_2 = M$, then both equivalence $\equiv$ and k-equivalence $\equiv_k$ are equivalence relations that they obey the reflexive, symmetric and transitive laws. We denote the partition corresponding to $\equiv$ by $Q/\equiv$ and the partition corresponding to $\equiv_k$ by $Q/\equiv_k$.

In our next step we are giving the theorem which will prove the two expansions $\delta^\square$ and $\lambda'$.

**Theorem 3.1:** Let $M = (Q, \Sigma, \Delta, \delta, \lambda)$ be a Linguistic Finite Automata. Then $\delta^\square$ and $\lambda^\square$.

1. $\delta = \delta'|Q \times \Sigma \times Q$. $\qquad\qquad (4)$
2. $\lambda = \lambda'|Q \times \Sigma \times \Delta$. $\qquad\qquad (5)$

Proof: Let $x \in \Sigma$, Then $\delta^\square(x) = \delta^\square(\varepsilon x) = \delta(\varepsilon) \circ \delta(x) = E_{n \times n} \circ \delta(x) = \delta(x)$ (4).

Let $q \in Q, x \in \Sigma, y \in \Delta, \lambda^\square(q, x, y) = \lambda^\square(q, \varepsilon x, \varepsilon x) = \lambda^\square(q, \varepsilon, \varepsilon) \wedge \{\vee[\delta^\square(q, \varepsilon, r) \wedge \lambda(r, x, y)| r \in Q]\} = \vee[\delta^\square(q, \varepsilon, r) \wedge \lambda(r, x, y) | r \in Q] = \delta^\square(q, \varepsilon, q) \wedge \lambda(q, x, y) = \lambda(q, x, y)$.

We shall prove that the length of input word must be the same as that of output word through mathematical induction.

## 4. Conclusion

We have introduced an alternative interpretation of Linguistic variables to define our proposed Linguistic Finite Automata along with its equivalence relations and two most important expansions $\delta^\square$ and $\lambda^\square$. Compared with the old model of Finite Automata holding all the conditions our modified model Linguistic Finite Automata is able to handle the Linguistic data and transitions which is not possible for the general kind of Finite Machine. Here the Linguistic quantifiers are treated in the same way being modeled as linguistic descriptions of probability. As such they encode second order density functions and can be viewed as an alternative form of imprecise probability.

**Acknowledgments**

Supriya Raheja Author thanks Dr. Reena Dadhich who provided invaluable comments in. Author also thanks Dr. Smita Rajpal for her inputs in the conceptualization of the paper and her support throughout for technical discussions.

**Supriya Raheja**, ITM University, pursuing her PhD in Computer Science from Banasthali University. She had done her engineering from Hindu college of Engineering, Sonepat and masters from Guru Jambeshwar University of Science and Technology, Hisar. She is working as a Reviewer/Committee member of various International Journals and Conferences. Her total Research publications are seven.

**Dr. Reena Dadhich** is presently working as an Associate Professor and Head of the Department of Master of Computer Applications at Engineering College Ajmer, India. She received her Ph.D. (Computer Sc.) and M.Sc. (Computer Sc.) degree from Banasthali University, India. Her research interests are Algorithm Analysis & Design, Wireless Ad-Hoc Networks and Software Testing. She has more than 12 years of teaching experience. She is working as an Editorial Board Member / Reviewer/Committee member of various International Journals and Conferences. She has written many research papers and books.

**Dr. Smita Rajpal**, ITM University, completed her PhD in Computer Engineering. She has a total work experience of 11 years. She is specialized in TOC, Compiler Design, Soft Computing and RDBMS. She is a Java certified professional. She is working as an Editorial Board Member / Reviewer/Committee member of various International Journals and Conferences. She is an active member of IEEE. Her biography is a part of Marquis who's who in the world, 2010.Her total Research publications are 20 and book chapter's-5.She has published three books.